\documentstyle[prb,eqsecnum,aps]{revtex}

\font\bba=msbm10 scaled 1080
\font\bbb=msbm8 
\font\bbc=msbm6 
\newfam\bbfam
\textfont\bbfam=\bba
\scriptfont\bbfam=\bbb
\scriptscriptfont\bbfam=\bbc

\title
{Sine-Gordon theory for the equation of state of classical hard-core Coulomb
systems.\\
 II. High-temperature expansion.}

\author{Jean-Luc Raimbault \thanks{Electronic mail:
raimbaut@lpmi.polytechnique.fr}}
\address{LPTP - CNRS (UMR 7648) \\
\'Ecole Polytechnique \\
91128 Palaiseau Cedex, France}      
\author{Jean-Michel Caillol \thanks{Electronic mail: 
Jean-Michel.Caillol@labomath.univ-orleans.fr}}
\address{MAPMO - CNRS (UMR 6628)\\
         D\'epartement de Math\'ematiques \\
         Universit\'e d'Orl\'eans, BP 6759\\
         45067 Orl\'eans cedex 2, France}         

\begin{document}
\draft
\date{\today}
\maketitle
\begin{abstract}
We perform a high-temperature expansion of the grand potential of the
restrictive primitive model of electrolytes
in the frame of the extended sine-Gordon theory exposed in the companion paper.
We recover a result already obtained by Stell an Lebowitz
(J. Chem. Phys. {\bf 49}, 3706 (1968)) by means of diagrammatic expansions.
\end{abstract}
\section{Introduction}
\label{intro}
In the first part of this work (hereafter referred as I\cite{I}) 
we have established an
exact field theoretical representation of the statistical mechanics of the
restricted primitive model of electrolytes (RPM), i.e. an equimolar mixture of
positively and negatively charged hard spheres of the same diameter $\sigma$.
Although the generalization to assymmetric models is certainly possible we have
limited ourselves to the case where the cations and the anions bear opposite
charges $\pm e$. By applying the Kac-Hubbard-Stratonovich-Siegert-Edwards
\cite{Kac,Hubbard,Hubbard2,Stratonovich,Siegert,Edwards}
transform to the Coulomb
interaction we have obtained an exact new expression of the action 
${\cal S}[\phi]$
very similar to the well-known Sine-Gordon action to which it reduces in the
limit of vahishing hard-core diameters.\cite{Samuel,Martin} 

By contrast with recent works in that direction\cite{Brillantov,Orland} our
formalism incorporates correctly the hard-core effects and our Sine-Gordon
action is well defined as a consequence of a smearing of the charge of the ions
which regularizes the Coulomb interactions at short distances. In the companion
paper we have applied our formalism to a derivation of the equation of state of
the RPM in the low activity ($z$) regime.
 The old results obtained in the frame of the
Mayer diagrammatic theory\cite{Mayer,Haga,Fisher} 
and recently rediscovered by Netz and Orland\cite{Orland} are confirmed and our
derivation precises the domain of validity of the (approximate) field
theoretical formulation of Netz and Orland. 

In the present paper we show that the formalism developped in\ I also enables a
derivation of the equation of state of the RPM in the high-temperature regime
and we recover the results obtained by Stell and Lebowitz\cite{Stell} in the
frame of the $\gamma$-ordering theory of liquids.\cite{Stell2,Hemmer,Hauge} The
latter is a sophisticated diagrammatic perturbation theory in which the "small"
parameter is $\gamma \equiv \beta e^2$ ($\beta$ inverse temperature), otherwise
the density is arbitrary. Note that a derivation of the 
equation of state of the RPM at high density is
clearly out of the scope of the theory of
Netz and Orland in which hard-core effects are incorporated only at the level of
the second virial coefficient. 

Our paper is organized as follows. In Sec.\ \ref{action} we reorganize the
Sine-Gordon action ${\cal S}[\phi]$ of the RPM as an expansion in powers of
$\gamma$. A $\gamma$-ordering of the grand-potential $\omega_{RPM}$
is then obtained by making use
of the cumulant theorem.\cite{Ma} Note that we thus make use of
the same technics as in I to obtain the $z$-ordering of  $\omega_{RPM}$
which confers an aesthetic unity to our approach. 
In Sec.\ \ref{omega-5/2} we derive an
expression of  $\omega_{RPM}$ valid at order $\gamma^{5/2}$. Moreover we show
that our expression for $\omega_{RPM}$ 
is independent of the precise nature of the smearing (i.e.
independent of the smearing diameter $a$, $0<a\leq\sigma$) at this order. This
result completes the independence of the $z$-expansion of $\omega_{RPM}$ upon
$a$ which
was proved in the companion paper. Admitting this independence at higher orders
in $\gamma$ greatly facilitates the algebra of Sec.\ \ref{omega-7/2} where
$\omega_{RPM}$ is obtained at order $\gamma^{7/2}$ by performing the
calculations with $a\to 0$. A Legendre transform then gives an expression of
the free energy of the RPM at order $\gamma^{7/2}$ which coincides with the old
result of Stell and Lebowitz.\cite{Stell} We conclude in Sec.\ \ref{conclusion} 
\section{$\gamma$-Ordering of the grand-potential}
\label{action}
Our starting point will be the expression of
the grand potential per unit volume
$ \omega_{RPM}\left(\beta,\nu\right)
\equiv - \ln\left[\Xi_{RPM}\left(\beta,\nu,L^3\right)
\right]/L^3$ ( $\Xi_{RPM}\left(\beta,\nu,L^3\right)$ grand-partition function)
of the RPM, which was derived in\ I in the frame of a regularized
Sine-Gordon \cite{Edwards,Kac,Siegert,Stratonovich,Samuel,Martin}
 theory.
Here 
$\nu=\beta \mu$ ($\beta$ inverse temperature),
 where $\mu = \mu_{\pm}$ denotes the chemical potentials of the
anions and the cations which can be chosen equal for simplicity.\cite{Martin}
 Moreover, we implicitly assume that the ions
are enclosed in a cubic box of side $L$, with periodic boundary conditions.
Recall that ({\em cf.} Eq\ (3.1) of ref.\ I).
\begin{mathletters}
\label{omega_RPM}
\begin{eqnarray}
\label{omega_RPM_a}
 \omega_{RPM}(\beta,\nu) &=&  \omega_{HS}(\nu_0)  + \omega_{1}
+ \Delta \omega \; , \\
\label{omega_RPM_b}
 \omega_{1} &=& \frac{1}{2}
\int \frac{d^3\vec{q}}{(2\pi)^3} \; \ln \left( 1
+ \gamma \rho_0 \tilde{W}_\tau(q) \right) \;  \\
\label{omega_RPM_c}
L^3 \Delta \omega &=& -\ln \langle
\exp \left( -{\cal H}\left[ \phi \right] \right)
\rangle_{X_{\tau}} \; .
\end{eqnarray}
\end{mathletters}
In eq.\ (\ref{omega_RPM_a}) $ \omega_{HS}(\nu_0)$ is the grand potential 
(per unit volume) of a
fluid of neutral hard spheres at the chemical potential $\nu_{0}=\nu+\ln(2)$.
 In Eq.\ (\ref{omega_RPM_b})
 we have denoted by $\rho_{0}\equiv
\rho_{HS}(\nu_0)=-\partial
 \omega_{HS}/\partial \nu_{0}$ the number density of this reference system.
In Eq.\
(\ref{omega_RPM_b}) $\tilde{W}_\tau(q)$ is the Fourier transform of the
 electrostatic interaction ${W}_\tau(r)$ between two balls of diameter
$a$ distant of $r$, each of unit charge and
with an uniform surfacic charge density $1/\pi a^2$.
The smearing of the charge  over the surface of a sphere of
diameter $a$ smaller than the diameter $\sigma$ of the ions 
was introduced in\ I in order to regularize the Sine-Gordon transform. 
Since 
\begin{equation}
\tilde{W}_\tau(q) = \frac{4 \pi}{q^2} 
\left(\frac{ \sin (qa/2)}{qa/2}\right)^2 \; ,
\end{equation}
the expression\ (\ref{omega_RPM_b}) of the contribution
$\omega_{1}$ is convergent for $a \neq 0$ as a consequence of the smearing. 
It must be stressed that the expression\ (\ref{omega_RPM}) of
$\omega_{RPM}$ is exact and  does not
depend on the value of the smearing diameter $a$ 
(as long as $0 <a \leq \sigma$).

The last contribution\ (\ref{omega_RPM_c}) to the grand potential involves the
effective hamiltonian
\begin{mathletters}
\label{uphi}
\begin{eqnarray}
{\cal H}[\phi] &=& U[\phi]-U_0[\phi] \; , \\
U_0[\phi]&\equiv&  \frac{\rho_0 \gamma}{2} \int d^3\vec{r}_1\; \phi^2(1) \; , \\
U[\phi]&\equiv&
 -\sum_{n=1}^{\infty} \frac{\rho_0^n}{n!}
\int d^3\vec{r}_1 \cdots d^3\vec{r}_n \;
h_0^{(n)} (1, \cdots, n)
\prod_{i=1}^{n}\left[\exp(\gamma/a)
\cos\left( \sqrt{\gamma}\phi(i)\right) - 1 \right]\; ,
\end{eqnarray} 
\end{mathletters}
where $\gamma = \beta e^2$ ($\pm e$ charges of the ions) and where
the $h_0 ^{(n)}(1, \cdots, n)$ are the connected n-body correlation functions of
the reference hard sphere fluid at the density $\rho_0$. These functions, which
are closely related to the Ursell distribution functions,
 \cite{Percus} are solution of a
hierarchy of integral equations which is briefly rewied
in Appendix\ \ref{appBa}. 
In Eq.\ (\ref{omega_RPM_c}) the
brackets $\langle \ldots \rangle_{X_{\tau}}$ denote a Gaussian average
over the fluctuations of a real scalar field $\phi$ 
with covariance $X_{\tau}$; the latter 
 is a screened potential related
to the Coulombic potential $W_{\tau}$ by the relation 
\begin{equation}
\tilde{X}_{\tau}(q) =
\frac{\tilde{W}_{\tau}(q) }
{1 +  \gamma \rho_0 \tilde{W}_{\tau}(q)} \; .  
\end{equation}

In the companion paper\ I, we have given a low-fugacity expansion of
$\omega_{RPM}$, here we rather consider the behavior of the grand potential at
high temperatures and arbitrary chemical potential $\nu$. 
In that aim we first
reorganize the Sine-Gordon Hamiltonian ${\cal H}\left[ \phi \right]$
as an expansion in the (small) parameter $\gamma$ of Stell and
Lebowitz\cite{Stell}. It is easily established 
from Eqs.\ (\ref{uphi}) that
\begin{equation}
\label{cumul1}
{\cal H}[\phi] = -\gamma \rho_0 L^3/a +
\sum_{N=2}^{\infty}\gamma^N\; W_N
\; ,
\end{equation}
where
\begin{mathletters}
\label{WN}
\begin{eqnarray}
W_N&=& \sum_{n=1}^{N}\; I_N^{(n)} \; , \\
\label{IN}
I_N^{(n)}& =& -\frac{\rho_0^n}{n!}\int d^3\vec{r}_1 \ldots d^3\vec{r}_n \;
 h_0 ^{(n)}
(1, \cdots, n) 
{\sum}^{*}{\cal F}^{(p_1)}(\phi_1) \ldots {\cal F}^{(p_n)}(\phi_n)
\; , \\ 
{\cal F}^{(p)}(\phi) &= &\sum_{s=0}^{p} \frac{(-1)^s}{(2s)!}
 \frac{a^{s-p}}{(p-s)!}
\phi^{2s} \; .
\end{eqnarray}
\end{mathletters}
The summations in the r.h.s of Eq.\ (\ref{IN}) are restricted to all 
the integers $p_i \geq 1$ ($i=1,\ldots,n)$ such that $\sum _{i=1}^{n} p_{i}=N$,
which was denoted by the symbol $\sum^*$. 
Later, we shall need  explicit expressions of $W_2$ and $W_3$; they are given
in Appendix\ \ref{appW}.

In order to obtain a
$\gamma$-expansion of $\Delta \omega$ we first apply the cumulant
theorem\cite{Ma} to Eq.\ (\ref{omega_RPM_c}) which leads to
\begin{equation}
\label{cumuH}
\ln\langle \exp\left(-{\cal H}\left[\phi\right]\right)\rangle_{X_{\tau}}=
\frac{\gamma \rho_0 L^3}{a} - \langle{\cal
H}\left[\phi\right]\rangle_{X_{\tau ,c}} +
\frac{1}{2}  \langle{\cal
H}^2\left[\phi\right]\rangle_{X_{\tau ,c}} + \ldots \; ,
\end{equation}
and then re-order the terms in the r.h.s 
with the help of Eq.\ (\ref{cumul1}) which yields 
\begin{eqnarray}
\label{Deltaomega}
L^3 \Delta \omega \equiv - \ln\langle \exp\left(-{\cal
H}\left[\phi\right]\right)\rangle_{X_{\tau}}&=&
-\frac{\gamma \rho_0 L^3}{a}
+\gamma^2 \langle W_2\rangle_{X_{\tau,c}} +
\gamma^3\langle W_3\rangle_{X_{\tau,c}} \nonumber \\
&+&
\gamma^4 \left( \langle W_4\rangle_{X_{\tau,c}} -\frac{1}{2} \;
\langle W_2^2\rangle_{X_{\tau,c}} \right) +
\ldots \; .
\end{eqnarray}
Note that  the subscript "c" in Eqs.\ (\ref{cumuH}), (\ref{Deltaomega}) 
denotes a cumulant average, for instance we have $\langle
W_N\rangle_{X_{\tau,c}}  \equiv\langle
W_N\rangle_{X_{\tau}}$ and $\langle
W_N^2\rangle_{X_{\tau,c}} \equiv \langle
W_N^2\rangle_{X_{\tau}} - \langle W_N\rangle_{X_{\tau}}^2$.
Gathering Eqs.\ (\ref{omega_RPM}), (\ref{Deltaomega}) we can write finally
\begin{mathletters}
\begin{eqnarray}
\label{om}
\omega_{RPM}\left(\beta,\nu\right)&=&\omega_{HS}(\nu_0)  + \delta \omega_{1}
+ \Delta \overline{\omega} \; ,  \\
\delta \omega_{1}&=& \omega_{1} -\frac{\rho_0 \gamma}{a} \; , \\
\label{om_c}
L^3\Delta \overline{\omega} &=& \gamma^2 \langle W_2\rangle_{X_{\tau,c}} +
\gamma^3\langle W_3\rangle_{X_{\tau,c}}   -\frac{\gamma^4}{2}\;
\langle W_2^2\rangle_{X_{\tau,c}} + {\cal O}(\gamma^4)
\ldots \; .
\end{eqnarray}
\end{mathletters}
Eq.\ (\ref{om_c}) needs some comments. The first order cumulants $\langle W_N
\rangle_{X_{\tau,c}}$ which enter the equation 
originate from the $\gamma$-ordering of $\langle {\cal H}[\phi]
\rangle_{X_{\tau,c}}$ which is a regular function of $\gamma$ in the limit $\gamma
\to 0$. Indeed  $\langle {\cal H}[\phi]
\rangle_{X_{\tau,c}}$ is obtained by applying Wick's
theorem\cite{Ma,Binney,Parisi} to Eq.\ (\ref{uphi}) and, 
since in the limit  $\gamma \to 0$ we have $X_{\tau} \to W_{\tau}$ 
 the resulting expression of  $\langle
{\cal H}[\phi] \rangle_{X_{\tau,c}}$ will involve integrals of 
products of long-range Coulombic
terms by the $n$-body correlations $h_0^{(n)}$ of the
hard sphere reference fluid which are assumed to be short-ranged and make the
integrals convergent. Therefore the first order cumulants $\langle W_N
\rangle_{X_{\tau,c}}$ are  regular in the limit $\gamma \to 0$
 but it could be that
some higher order cumulants are singular.
 For instance, it will be shown in Sec.\ \ref{omega-7/2} that 
$\langle W_2^2\rangle_{X_{\tau,c}}$ diverges as $\gamma^{-1/2}$ in the limit
$\gamma \to 0$. In writing Eq.\ (\ref{om_c}) we have assumed that,
 at this order,
higher order cumulants do not contribute to $\Delta \overline{\omega}$,
which is, admittedly, difficult to prove rigorously.

We stress again that, although the variable "$a$" appears explicitly in Eqs.\ (\ref{om}) as well as in
the expressions\ (\ref{WN}) of the $W_N$, the grand-potential $\omega_{RPM}$
does not depend upon "$a$", $\forall a \in]0,\sigma]$. Moreover, in
a perturbation
scheme such as the $\gamma$-ordering this independence should certainly
be satisfied for
each term of the expansion.
 We have checked this point up to order $\gamma^{5/2}$. This result,
which is discussed in details in the next section, completes the results of
paper\ I where we have checked that each term of the low activity expansion of
$\omega_{RPM}$ was actually independent of $a$.

\section{$\gamma$-expansion  of  $\omega_{RPM}(\beta,\nu)$
 up to order $\gamma^{5/2}$}
\label{omega-5/2}
Since, in Eq.\ (\ref{om}), $\omega_{HS}(\nu_0)$ does not depend upon the
temperature, we are left with the problem of finding a $\gamma$-expansion of the
terms $\delta \omega_{1}$ and $\Delta \overline{\omega}$ 
of Eq.\ (\ref{om}) up to order ${\cal
O}(\gamma^{5/2})$. We have shown in\
I that, for a given $a$, $\omega_{1}$ was a function of the sole parameter
$(\rho_0 \gamma)^{1/2}$; therefore the expansions of
$\omega_{1}$ in powers of $(\rho_0 \gamma)^{1/2}$ 
given in I are valid either at low activities (or equivalently at
low $\rho_0$) or
at high temperatures (i.e. at low $\gamma$),
which is the case that we consider here.
Recall that (cf. Eq.\ (3.6) of\ I)
\begin{equation}
\label{omega1}
\delta \omega_1 = -\frac{2}{3\pi} \left(\pi \rho_0 \gamma \right)^{3/2}+
\frac{7}{15 \pi} \left(\pi \rho_0 \gamma \right)^2 a -
\frac{1}{3\pi} \left( \pi \rho_0 \gamma \right)^{5/2} a^2 +{\cal O}\left(
\left( \gamma a\right)^3\right) 
\; .
\end{equation}
Note that $\delta \omega_1$ is regular in the limit $a \to 0$ and that
$\lim_{a \to 0} \delta \omega_1=-2\left(\pi \rho_0 \gamma
\right)^{3/2}/3\pi$ which is reminiscent of the well-known
Debye-H\"uckel result\cite{Debye,Hansen}.

On the other hand, up to order ${\cal
O}(\gamma^{5/2})$   we have, from Eq.\ (\ref{om_c})
\begin{equation}
L^3 \Delta \overline{\omega} = \gamma^2 \langle W_2\rangle_{X_{\tau,c}}+ {\cal
O}(\gamma^3) \; .
\end{equation} 
The expression of $W_2$ is given in Appendix \ref{appW}. 
Taking the Gaussian average of Eq.\ (\ref{WW2}) on finds 

\begin{eqnarray}
\label{W2}
\langle W_2 \rangle_{X_{\tau}} &=& -\rho^{'}_0 \frac{L^3}{2a^2} + \frac{
\rho^{'}_0}{2a} \int d^3\vec{r}_1 \; \langle \phi^2(1) \rangle_{X_{\tau}}
-\frac{\rho_0}{4!} \int d^3\vec{r}_1 \; \langle \phi^4(1) \rangle_{X_{\tau}}
\nonumber \\
& -&\frac{\rho_0^2}{8} \int d^3\vec{r}_1 d^3\vec{r}_2 \; h_0^{(2)}(1,2) \langle \phi^2(1)
\phi^2(2)
\rangle_{X_{\tau}} \; , 
\end{eqnarray} 
where $\rho^{'}_0$ is the derivative of the number density of the hard sphere
reference fluid $\rho_0$ with respect to the
chemical potential $\nu_0$.
 The Gaussian averages which enter Eq.\ (\ref{W2}) are computed with
the help of Wick's theorem (details are given in Appendix\ \ref{appWi}).
Inserting Eqs.\ (C1), (C2), (C5) in Eq.\ (\ref{W2}) yields

\begin{equation}
\label{W2sL3}
\gamma^2\langle W_2 \rangle_{X_{\tau}}L^{-3} =
\rho^{'}_0 \gamma^2 \left( -\frac{1}{2a^2}+\frac{X_{\tau}(0)}{2a}
-\frac{X_{\tau}^2(0)}{8} \right) -\frac{\gamma^2\rho_0^2}{4}
\int d^3\vec{r} \; h_0^{(2)}(r)X_{\tau}^2(r) \; ,
\end{equation}
where $h_0^{(2)}(r_{12}) \equiv h_0^{(2)}(12)$ is the usual pair correlation
function in the limit of an infinite homogeneous and isotropic
system. Note that, in the core, i.e. for
$0\leq r \leq\sigma$, $h_0^{(2)}(r)=-1$. In order to derive Eq.\ (\ref{W2sL3})
we have also made use of the compressibility sum rule for the reference hard
sphere fluid\cite{Hansen}, i.e.
\begin{equation}
\label{compres}
\rho_0^2 \tilde{h}_0^{(2)} (0)=\rho^{'}_0-\rho_0 \; ,
\end{equation}
where $\tilde{h}_0^{(2)}(q)$ is the $3D$ Fourier transform of $h_0^{(2)}(r)$.
We note that 
the screened potential $X_{\tau}(r)$ is a function which depends
on the smearing diameter $a$ and therefore, {\em a priori}, $\langle W_2
\rangle_{X_{\tau}} $ as given by Eq.\ (\ref{W2sL3}) depends upon $a$ in a
complicated manner. It is
this dependance that we want to study in detail now.

We have shown in the companion paper that, for a given smearing diameter $a$,
the screened potential $X_{\tau}(r)$
was a function of the sole variable 
\begin{equation}
\label{kappa}
\kappa_0 \equiv 2(\pi \rho_0 \gamma)^{1/2} \; .
\end{equation}
Note that $\kappa_0$ is not, strictly speaking, the inverse Debye length of the
system since it involves the density $\rho_0$ of the reference system which is
not equal, a priori, to the density $\rho$ of the RPM.
We have shown that,
for $r=0$ ({\em cf} Eq.\ (3.8) of I) 
\begin{equation}
X_{\tau}(0)=\frac{2}{a} 
-2(\pi\rho_0\gamma)^{1/2} + \frac{28}{15} (\pi\rho_0\gamma)a -\frac{5}{3}
(\pi\rho_0\gamma)^{3/2} a^2 +{\cal O}(\gamma^2 a^3) \; ,
\end{equation}
from which it follows that the first term of the r.h.s. of Eq.\ (\ref{W2sL3})
is regular as $a\to 0$ and reads as
\begin{equation}
\label{firstterm}
\rho^{'}_0 \gamma^2 \left( -\frac{1}{2a^2}+\frac{X_{\tau}(0)}{2a}
-\frac{X_{\tau}^2(0)}{8} \right) = -\frac{1}{2}
\pi\rho_0\rho^{'}_0\gamma^3 + {\cal O} (\gamma^{7/2}a) \; .
\end{equation}
For $r\neq 0$, $X_{\tau}(r)$ is a piecewise function defined as :
\begin{mathletters}
\label{X(r)}
\begin{eqnarray}
X_{\tau}^{>}(r) &=& q^2(\kappa_0a/2)
 \frac{\exp(-\kappa_0r)}{r} +{\cal O}(\gamma a) \text{ for }
r\geq a \; , \\
X_{\tau}^{<}(r) 
                &=& \frac{2}{a}\left(1-\frac{r}{2a}\right)
 -\kappa_0 + {\cal O}(\gamma a) \text{ for }
r\leq a \; ,
\end{eqnarray}
\end{mathletters}
where $q(x)\equiv \sinh(x/2)/(x/2)$. Note by passing that Eqs.\ (\ref{X(r)}) 
imply that 
\begin{equation}
\label{X-tau}
\lim_{a\to 0} X_{\tau}(r)\equiv y(r) =\exp(-\kappa_0 r)/r \; \; 
\forall r.
\end{equation}
Inserting the expressions\ (\ref{X(r)}) of $X_{\tau}(r)$ 
in the integral of the r.h.s. of Eq.\
(\ref{W2sL3}) and noting that $h_0^{(2)}(r)=-1$ for $0\leq r \leq a \leq
\sigma$ yields
\begin{eqnarray}
\label{olp}
-\frac{\rho_0^2 \gamma^2}{4}\int d^3\vec{r} \; h_0^{(2)}(r)X_{\tau}^2(r)&=&
-\frac{\rho_0^2 \gamma^2}{4}\int d^3\vec{r} \; h_0^{(2)}(r)X_{\tau}^{>2}(r)
\nonumber \\
&-&\frac{7}{15\pi}  (\pi\rho_0 \gamma)^2 a + \frac{1}{3\pi} (\pi \rho_0
\gamma)^{5/2} a^2 + {\cal O}\left(\left(\gamma a\right)^{3}\right) \; ,
\end{eqnarray}
where the terms proportional to $a$ and $a^2$ in the r.h.s.
 of Eq.\ (\ref{olp})
originate from the integration of $X_{\tau}^{<2}(r)-X_{\tau}^{>2}(r)$ in the
core (where $h_0^{(2)}(r) =-1$).
Moreover, note that, in Eq.\ (\ref{olp}) we can safely expand $X_{\tau}^{>}(r)$
in powers of
$\kappa_0$ since $h_0^{(2)}$ is a short range function of $r$. Performing this
expansion and gathering Eqs.\ (\ref{omega1}), (\ref{firstterm}) and
 (\ref{olp}) we obtain finally, at order $\gamma^{5/2}$
\begin{eqnarray}
\label{omega5/2}
\omega_{RPM}(\beta,\nu) &=& \omega_{HS}(\nu_0) -\frac{2}{3\pi} \left(\pi \rho_0
\gamma \right)^{3/2} - \frac{\left(\rho_0 \gamma \right)^{2}}{4}
\int d^3\vec{r} \; \frac{h_0^{(2)}(r)}{r^2} \nonumber \\
&+& \pi^{1/2} \left( \rho_0 \gamma \right)^{5/2} \int d^3\vec{r} \;
\frac{h_0^{(2)}(r)}{r} +{\cal O}(\gamma ^{3}) \; .
\end{eqnarray}
Due to a complex mechanism of compensations,
 the expression\ (\ref{omega5/2}) of
the grand potential at order $\gamma^{5/2}$ is, as anticipated, indeed 
independent of the value of the smearing diameter $a$. 
\section{$\gamma$-expansion  of  the grand potential and the free energy
 up to order $\gamma^{7/2}$}
\label{omega-7/2}
In this section we
obtain the expansion of $\omega_{RPM}(\beta,\nu)$ up to order
$\gamma^{7/2}$. In order to make the derivations more tractable,
we shall assume that the terms of order greater than $\gamma^{5/2}$ 
in the $\gamma$-expansion of
$\omega_{RPM}(\beta,\nu)$ are still independent of $a$. Moreover we
shall consider
the limit $a\to 0$ of Eq.\ (\ref{om}) which allows us to replace
$X_{\tau}(r)$ by its limiting
expression $y(r) \equiv \exp(-\kappa_0r)/r$ and greatly facilitates the
derivations. Therefore one has

\begin{mathletters}
\begin{eqnarray}
\label{om_bis}
\omega_{RPM}\left(\beta,\nu\right)&=&
\lim_{a\to0} \left(\omega_{HS}(\nu_0)  + \delta \omega_{1}
+ \Delta \overline{\omega} \right) \; , \\
\label{om_ter}
L^3\Delta \overline{\omega} &=& \gamma^2 \langle W_2\rangle_{X_{\tau,c}} +
\gamma^3\langle W_3\rangle_{X_{\tau,c}} -
\frac{\gamma^4 }{2} \;\langle W_2^2\rangle_{X_{\tau,c}} +
{\cal O}({\gamma^4})\; .
\end{eqnarray}
\end{mathletters}
The limit of $\delta \omega_{1}$ for $a \to 0$  is easily obtained
 from Eq.\ (\ref{omega1}) and reads as 
\begin{equation}
\lim_{a \to 0} \delta \omega_{1} = -\frac{\kappa_0^3}{12 \pi} =
- \frac{2}{3\pi} \left(\pi \rho_0 \gamma \right) ^{3/2} \; .
\end{equation}
The expression  of the term $\langle W_2\rangle_{X_{\tau,c}}$ in the r.h.s of
Eq.\ (\ref{om_ter}) has already been derived in Sec.\ \ref{omega-5/2}. We take
the limit $a \to 0$ of Eq.\ (\ref{W2sL3}) which gives

\begin{eqnarray}
\lim_{a\to 0}\gamma^2 \langle W_2\rangle_{X_{\tau,c}}L^{-3} &=&
-\frac{\pi \rho_0 \rho_0^{'} \gamma^3}{2}
-\frac{\gamma^2 \rho_0^2}{4} \int d^3\vec{r} \; h^{(2)}_0(r) \frac{\exp(-2
\kappa_0 r)}{r^2} \nonumber \; , \\
&=& -\frac{\rho_0^2 \gamma^2}{4}  \int d^3\vec{r} \; h^{(2)}_0(r) \frac{1}{r^2}
+\pi^{1/2}\left(\rho_0 \gamma \right)^{5/2}  
\int d^3\vec{r} \; h^{(2)}_0(r) \frac{1}{r} +
\pi \rho_0 \gamma^3\left( 2 \rho_0 -\frac{5}{2} \rho_0^{'} \right) \; \nonumber
\\
&+& \frac{8\pi^{3/2}}{3} \left(\rho_0 \gamma \right)^{7/2}
\int d^3\vec{r} \; h^{(2)}_0(r) r + {\cal O}(\gamma^{4}) \; ,
\end{eqnarray}
where we have expanded the Yukawa potential in powers of $\kappa_0$, taking
advantage of the assumed short-range behaviour of $h_0^{(2)}(r)$ and also
made use of the compressibility sum rule\ (\ref{compres}).

The expression of $W_3$ is given in Appendix\ \ref{appW}. Taking the Gaussian
average of Eq.\ (\ref{WW3}) and then making use of Wick's theorem and of the
relation\ (\ref{ur2}) of
Appendix\ \ref{appBa} yields 

\begin{eqnarray}
\label{pil}
\langle W_3\rangle_{X_{\tau,c}}L^{-3} &=&
w_3(a) + \int d^3\vec{r}\; X_{\tau}^2(r) \frac{\partial}{\partial \nu_0}
\left(\rho_0^2 h_0^{(2)}(r) \right) \left( -\frac{1}{4a} +\frac{X_{\tau}(0)}
{8} \right) \; , \nonumber \\ 
&+& \frac{\rho_0^3}{6} \int d^3 \vec{r}_2 d^3 \vec{r}_3 \; h_0^{(3)}(1,2,3)
 X_{\tau}(1,2) X_{\tau}(1,3)  X_{\tau}(2,3)\; \nonumber \\
 \lim_{a \to 0} w_3(a)&=& 
 -\frac{\rho_0^{''}\kappa_0^3}{48} \; .
\end{eqnarray}
We take now the limit $a \to 0$ of Eq.\ (\ref{pil}) and expand the resulting
expansion in powers of $\kappa_0$ which gives

\begin{eqnarray}
\lim_{a \to 0} \gamma^3 \langle W_3 \rangle_{X_{\tau,c}}L^{-3} &=&
\frac{\left(\rho_0 \gamma\right)^3}{6}
\int d^3 \vec{r}_2 d^3 \vec{r}_3 \; \frac{h_0^{(3)} (1,2,3)}{r_{12} r_{13}
r_{23}} \;
-\frac{\pi^{1/2} \rho_0^{1/2} \gamma^{7/2}}{4}
\int d^3\vec{r} \; \frac{\partial}{\partial \nu_0}
\left(\rho_0^2 h_0^{(2)}(r) \right) \frac{1}{r^2} 
\nonumber \\
&-& \pi^{1/2} \left(\rho_0 \gamma\right)^{7/2}
\int d^3 \vec{r}_2 d^3 \vec{r}_3 \; \frac{h_0^{(3)} (1,2,3)}
{r_{12} r_{23}}   + {\cal O}(\gamma^4) \; .
\end{eqnarray} 
Note that both cumulants $\langle W_2 \rangle_{X_{\tau}}$ and $\langle W_3
\rangle_{X_{\tau}}$ are regular in the limit $\gamma \to 0$ which confirms the
analysis made at the end of Sec. \ref{action}.

The analysis of the limit $a\to 0$ of the last contribution 
$-\gamma^4 \langle
W_2^2 \rangle_{X_{\tau,c}}/(2L^3)$ of the r.h.s of Eq.\ (\ref{om_ter})
is more delicate and postponed to Appendix\
\ref{appW22}. One finds that this term can be written as
\begin{eqnarray}
\lim_{a\to 0}\gamma^4 L^{-3} \langle W_2^2 \rangle_{X_{\tau,c}} &=&
\frac{\left(\rho_0 \gamma\right)^{7/2}}{4\pi^{1/2}}
\lim_{\kappa_0 \to 0} 
\int d^3 \vec{r}_2 d^3 \vec{r}_3 d^3 \vec{r}_4 \; 
h_0^{(2)}(1,2) h_0^{(2)}(3,4) \times \nonumber \\
&\times &\frac{\kappa_0 \exp\left(-\kappa_0 
\left(r_{12}+ r_{23}+ r_{34}+ r_{41}\right)\right)}
{r_{12} r_{23} r_{34} r_{41}}
+ {\cal O}(\gamma^4) \; .
\end{eqnarray}
By contrast with $\langle W_2 \rangle_{X_{\tau}}$ and $\langle W_3
\rangle_{X_{\tau}}$ the cumulant $\langle W_2^2 \rangle_{X_{\tau,c}}$ is
singular as $\gamma^{-1/2}$ in the limit $\gamma \to 0$. It is of course very
difficult to give a precise analysis of the singularities of higher order
cumulants.

Gathering the intermediate results we get our final expression for
the grand potential of the primitive model
up to order ${\cal O}(\gamma^{7/2})$.
\begin{mathletters}
\begin{eqnarray}
\label{omega-final}
\omega_{RPM}(\beta,\nu) &= &\omega_{HS}(\nu_0)+ \Delta \omega(\beta,\nu)
+{\cal O}(\gamma^{4}) \; ,\\
\Delta \omega(\beta,\nu)  &= & 
\sum_{i=1}^{9}\omega^{(i)}\; ,\\
\omega^{(1)} &=&
- \frac{2 }{3\pi} \left(\pi \rho_0 \gamma \right) ^{3/2} \; , \\
\omega^{(2)} &=& -\frac{\rho_0^2 \gamma ^2}{4} \int d^3 \vec{r}\;
\frac{h_0^{(2)}(r)}{r^2}\; , \\
\omega^{(3)} &=& \pi^{1/2} \left(\rho_0 \gamma \right)^{5/2} 
\int d^3 \vec{r}\; \frac{h_0^{(2)}(r)}{r}\; , \\
\omega^{(4)} &=& \pi \rho_0 \left( 2\rho_0-\frac{5}{2}\rho_0^{'}\right)
 \gamma^3 \; , \\
 \omega^{(5)} &=& \frac{\left( \rho_0 \gamma \right)^{3}}{6}
 \int d^3 \vec{r}_2 d^3 \vec{r}_3\  \frac{h_0^{(3)} (1,2,3)}{r_{12} r_{13} r_{23}}
 \; , \\
\omega^{(6)} &=& \frac{8}{3} \pi^{3/2} \left( \rho_0 \gamma \right)^{7/2}
\int d^3 \vec{r}\; h_0^{(2)}(r) \;r\; , \\
\omega^{(7)} &=&-\frac{\left(\pi \rho_0 \right)^{1/2}}{4} \gamma^{7/2}
\int d^3 \vec{r}\; \frac{\partial}{\partial \nu_0}
\left( \rho_0^2 h_0^{(2)}(r)\right)\frac{1}{r^2}  \; , \\
\omega^{(8)} &=& -\pi^{1/2} \left(\rho_0 \gamma\right)^{7/2}
\int d^3 \vec{r}_2 d^3 \vec{r}_3 \; \frac{h_0^{(3)} (1,2,3)}
{r_{12} r_{23}} \; , \\
\omega^{(9)} &=&-\frac{\left(\rho_0 \gamma\right)^{7/2}}{8\pi^{1/2}}
\lim_{\kappa_0 \to 0} 
\int d^3 \vec{r}_2 d^3 \vec{r}_3 d^3 \vec{r}_4 \; 
h_0^{(2)}(1,2) h_0^{(2)}(3,4) \times \nonumber \\
&\times & \frac{\kappa_0\exp\left(-\kappa_0 
\left(r_{12}+ r_{23}+ r_{34}+ r_{41}\right)\right)}
{r_{12} r_{23} r_{34} r_{41}} \; ; 
\end{eqnarray}
\end{mathletters}
where we have made a change of notation and redefined $\Delta \omega$.
The free energy $f_{RPM}(\beta,\rho)$ is defined as the Legendre transform of
$\omega_{RPM}(\beta,\nu)$ with respect to the chemical potential\cite{Hansen}, i.e.
\begin{equation}
\label{f}
f_{RPM}(\beta,\rho)= \omega_{RPM}(\beta,\nu) +\rho(\beta,\nu) \nu \;,
\end{equation}
where $\rho(\beta,\nu)$ is the density of the RPM which can be obtained by
differentiating $-\omega_{RPM}$ as given by Eq.\ (\ref{omega-final})
with respect to $\nu_0$. Therefore  we have
\begin{eqnarray}
\label{rho-RPM}
\rho &=& -\frac{\partial \omega_{RPM}}{\partial \nu_0}= \rho_0 +\Delta \rho \;
, \nonumber \\
\Delta \rho&=& \pi^{1/2} \rho_0^{1/2}\rho_0^{'} \gamma^{3/2} +
\frac{\gamma^2}{4} \int d^3 \vec{r} \; \frac{ \partial}{\partial \nu_0} 
\left( \rho_0^2 h_0^{(2)}(r) \right)\frac{1}{r^2} +{\cal O}(\gamma^{5/2}) \; . 
\end{eqnarray}
It should become clear in the sequel that an expansion of $\Delta \rho$ to order
$\gamma^{5/2}$ is sufficient for our purpose.
Let us denote by 
\begin{equation}
\label{deltanu}
\nu^*= \nu_0 +\Delta \nu \; ,
\end{equation}
the chemical potential of a fluid of hard spheres at the density $\rho$. Eq.\
(\ref{rho-RPM}) suggests that we search $\Delta \nu$ under the form
\begin{equation}
\label{dlnu}
\Delta \nu = \nu_{3/2} \gamma^{3/2} + \nu_2 \gamma^2 +{\cal O}(\gamma^{5/2})
\; ,
\end{equation}
In order to find the expressions of $\nu_{3/2}$ and $\nu_2$ we note that, by
definition of $\Delta \nu$ one has 
\begin{eqnarray}
\label{dlrho}
\rho &=& \rho_{HS}(\nu_0+\Delta \nu) \nonumber \\
&=& \rho_0 +\left(\nu_{3/2} \gamma^{3/2} + \nu_2 \gamma^2 \right) \rho_0^{'} +
{\cal O}(\gamma^{5/2}) \; .
\end{eqnarray}
Comparison of Eqs.\ (\ref{rho-RPM}) and (\ref{dlrho}) yields, by identification
\begin{mathletters}
\label{nu}
\begin{eqnarray}
\nu_{3/2} &=& \pi^{1/2} \rho_0^{1/2} \; , \\
\nu_2 &=& \frac{1}{4\rho_0^{'}} \int d^3 \vec{r} \frac{\partial}{\partial \nu_0}
\left(\rho_0^2 h_0^{(2)}(r) \right) \frac{1}{r^2} \; .
\end{eqnarray}
\end{mathletters}
We have now in hand all the ingredients to compute the free energy. Making use
of Eqs.\ (\ref{rho-RPM}), (\ref{deltanu}) and (\ref{dlnu})  one finds,
 at order $\gamma^{7/2}$
\begin{eqnarray}
\label{expref}
f_{RPM}(\beta,\rho) &=& \omega_{RPM}(\beta,\nu) +\rho \nu
 \nonumber \\
&=& \omega_{HS}(\nu^*-\Delta \nu) + \rho \nu +\Delta \omega \nonumber \\
&=& \omega_{HS}(\nu^*) + \rho \Delta \nu -\frac{1}{2} \rho_0^{'} 
(\Delta \nu)^2 +\rho \nu +\Delta \omega +{\cal O}(\gamma^4) \nonumber \\
&=& f_{HS}(\rho) +f^{(1)} +f^{(2)} + \Delta \omega +{\cal O}(\gamma^4) \; ,
\end{eqnarray} 
where $f_{HS}(\rho)\equiv \omega_{HS}(\nu^*) + \rho (\nu^* -\log2)$ 
is the free energy of the reference hard sphere fluid at the same
density $\rho$ than that of the RPM (note that the free energy of the
reference fluid and that of a fluid of identical hard spheres differ by a
mixing entropy term) and
\begin{mathletters}
\label{f12}
\begin{eqnarray}
f^{(1)} &=& -\frac{1}{2} \rho_0^{'} \nu_{3/2}^2 \gamma^3 \\
f^{(2)}&=& -\rho_0^{'} \nu_2 \nu_{3/2} \gamma^{7/2}
\end{eqnarray}
\end{mathletters}
In order to obtain a more transparent expression of $f_{RPM}$ we first remark
that
\begin{equation}
\label{remark1}
\omega^{(2)} + f^{(2)} = -\frac{\gamma^2}{4} \int d^3 \vec{r} \; \frac{1}{r^2} 
\left[ \rho_0^2 h_0^{(2)}(r) + \pi^{1/2} \rho_0^{1/2} \gamma^{3/2} 
\frac{\partial}{\partial \nu_0}\left(  \rho_0^2 h_0^{(2)}(r) \right) 
\right] \; ,
\end{equation}
and that the term in brackets  in the integral can be interpretated as a Taylor
expansion of the function $\rho_0^2 h_0^{(2)}(r)$. Since, from Eq.\
(\ref{dlnu}) $\Delta \nu = \pi^{1/2} \rho_0^{1/2} \gamma^{3/2} +{\cal
O}(\gamma^2)$, we can rewrite\ (\ref{remark1}) as
\begin{equation}
\label{omega2f2}
\omega^{(2)} + f^{(2)} = -\frac{\gamma^2 \rho^2}{4} \int d^3 \vec{r} \;
\frac{h^{*(2)}(r)}{r^2} +{\cal O}(\gamma^4) \; ,
\end{equation}
where we have denoted by $h^{*(2)}(r)$ the pair correlation function of the hard
spheres at the density $\rho$ ({\em i.e.} that of the RPM).

Our second remark is that the combination of several terms occuring in Eq.\
(\ref{expref}) give the usual Debye-H\"uckel free energy $-\kappa^3/12 \pi $.
Here $\kappa \equiv 2(\pi \rho \gamma)^{1/2}$ is the {\em true} Debye length of
the RPM which differs from $\kappa_0$ by terms of order $\gamma^{3/2}$. Indeed
one has
\begin{eqnarray}
\label{remark2}
-\frac{\kappa^3}{12 \pi} &\equiv &-\frac{2}{3} \pi^{1/2} \rho^{3/2} \gamma^{3/2}
\nonumber \\
&=& -\frac{2}{3} \pi^{1/2} \gamma^{3/2} \rho_0^{3/2}(1 + \frac{\Delta
\rho}{\rho_0})^{3/2} \nonumber \\
&=& -\frac{2}{3} \pi^{1/2} \rho_0^{3/2} \gamma^{3/2} - \pi^{1/2} \rho_0^{1/2}
\Delta \rho
\gamma^{3/2} +{\cal O}(\gamma^4) \nonumber \\
&=& \omega^{(1)} -\pi \rho_0 \rho_0^{'} \gamma^3 -
\frac{(\pi\rho_0)^{1/2}\gamma^{7/2}}{4}  
\int d^3 \vec{r} \frac{\partial}{\partial \nu_0}\left( \rho_0 h_0^{(2)}(r)
\right) \frac{1}{r^2} +{\cal O}(\gamma^4) \; , 
\end{eqnarray} 
By combining Eq.\ (\ref{remark2}) and the expressions of $\omega_1$, $\omega_4$,
$\omega_7$ one finds
\begin{eqnarray}
\label{remark22}
\omega^{(1)} + \omega^{(4)} +\omega^{(7)} + f^{(1)}
 &=& -\frac{\kappa^3}{12 \pi}+ 2 \pi \rho_0
\left( \rho_0 - \rho_0^{'} \right)\gamma^3 +{\cal O}(\gamma^4) \nonumber \\
&=& -\frac{\kappa^3}{12 \pi} + 2 \pi \rho
\left( \rho - \rho^{'} \right)\gamma^3 +{\cal O}(\gamma^4) \nonumber \\
&=& -\frac{\kappa^3}{12 \pi} -2 \pi \rho^3 \gamma^3 \int d^3 \vec{r}\; 
h^{*(2)}(r)  +{\cal O}(\gamma^4)\; ,
\end{eqnarray}
where the last line follows from the compressibility sum rule\ (\ref{compres})
and where we have noted that $\rho^{'}$ as well as  $\rho$ differ of
respectively $\rho_0^{'}$ and $\rho_0$ by terms of order $\gamma^{5/2}$. The
Legendre transform of $\omega_{RPM}$ is achieved by combining Eqs.\
(\ref{omega2f2}), (\ref{remark22}) and by
noting that the correlation functions $h_0^{(n)}$
 which enter terms of order $\gamma^{5/2}$
or higher can safely be replaced by the  $h^{*(n)}$ since
these functions differ by terms of order $\gamma^{3/2}$. Our final result for
$f_{RPM}$ is therefore
\begin{eqnarray}
\label{but}
f_{RPM}(\beta,\rho) &=& f_{HS}(\rho)  -\frac{\kappa^3}{12 \pi} -
\frac{\left(\rho \gamma \right)^{2}}{4} \int d^3 \vec{r}\;
\frac{h^{*(2)}(r)}{r^2} 
 + \pi^{1/2} \left( \rho \gamma \right)^{5/2}
 \int d^3 \vec{r}\; \frac{h^{*(2)}(r)}{r} \nonumber
\\
&-& 2 \pi \left( \rho \gamma\right)^{3} \int d^3 \vec{r}\; h^{*(2)}(r)
+\frac{\left(\rho \gamma \right)^{3}}{6}
\int d^3 \vec{r}_2 d^3\vec{r}_3 \;  \frac{h^{*(3)}(1,2,3)}{r_{12} r_{23} r_{31}}
\nonumber \\
&+& \frac{8}{3} \pi^{3/2} \left( \rho \gamma \right)^{7/2}
 \int d^3 \vec{r}\; r h^{*(2)}(r) -\pi^{1/2} \left( \rho \gamma \right)^{7/2}
 \int d^3 \vec{r}_2 d^3 \vec{r}_3 \;  
 \frac{h^{*(3)}(1,2,3)}{r_{12} r_{23}} \nonumber \\
 &-&\frac{\pi^{-1/2}}{8} \left( \rho \gamma \right)^{7/2}
 \lim_{\kappa \to 0} \int d^3 \vec{r}_2 d^3 \vec{r}_3 d^3 \vec{r}_4\;  
 h^{*(2)}(r_{12}) h^{*(2)}(r_{34}) \times \nonumber \\
&\times&\frac{\kappa \exp- \kappa\left(r_{12}
  + r_{23}
 +r_{34} + r_{41} \right)}{r_{12} r_{23} r_{34} r_{41} } +{\cal O}(\gamma^4)\; .
\end{eqnarray}
Some comments are in order. 
\begin{itemize}
\item{{\em (i)}} The expression\ (\ref{but}) coincides with that 
derived by Stell and Lebowitz by means of diagrammatic expansions in the frame
of the $\gamma$-ordering theory.\cite{Stell,Stell2,Hemmer,Hauge}
\item{{\em (ii)}} We have shown that (at least at order $\gamma^{5/2}$ and we
are enclined to admit this result at any orders) the expansion\ (\ref{but}) is
independent of the smearing diameter $a$. Therefore the choice $a \to 0$ could
have been made from the beginning, which extends the validity of\ (\ref{but}) to
other reference potentials than hard core repulsions (Lennard-Jones potentials
for instance). 
\end{itemize}
\section{conclusion}
\label{conclusion}
In this paper we have derived the $\gamma$-expansions of the grand-potential and
free energies of the RPM in the frame work of the exact field theoretical
formalism exposed in the companion paper.\cite{I} In I we had obtained
an activity expansion of the same quantities. In both cases of either high
temperatures (and arbitrary densities) or low fugacities (and arbitrary
temperatures) the perturbative expansions are obtained by a cumulant expansion of
the exact Sine-Gordon representation of the grand partition function $\Xi_{RPM}$
derived in the first paper. Of course one recovers in both cases expressions
derived years ago by means of sophisticated 
diagrammatic expansions involving more 
or less
complicated resummations of infinite series of graphs.\cite{Mayer,Haga,Stell}
The field theoretical representation of the grand partition function of a
theoretically important model such as the RPM (of course other models can be
considered too) can therefore be used to recover most of the results obtained in
the past by more traditional methods of classical statistical mechanics. 
However we have now at our disposal approximation schemes such that the
saddle point
approximation, the loop expansions etc. which have no real counterparts in the
conventional theory of liquids. Concerning the RPM, a mean field sine-Gordon
approximation of $\Xi_{RPM}$  is perhaps the correct starting point for a
full understanding of  ionic criticality wanted by M. E. Fisher.\cite{Fisher2}
Work in that direction is currently
in progress and will be reported elsewhere.

\appendix
\section{The hierarchy of Ursell functions }
\label{appBa}
Let us consider a classical fluid at inverse temperature $\beta$, chemical
potential $\mu$ in an external potential $\varphi(\vec{r})$. We denote by
$\Xi[\varphi]$ the grand partition function and by $z(\vec{r})
=\exp(-\beta \varphi(\vec{r})+\nu)/\lambda^3$ the activity ($\nu \equiv \beta
\mu$ and $\lambda \equiv$ de Broglie thermal wavelengh).
 In the absence of the external field $\varphi$, the $n$-body Ursell
(or connected) correlation functions are  defined as \cite{Percus,Hansen}
\begin{equation}
\label{ursell}
\rho_c^{(n)}(1, \ldots, n)
 = \left. \prod_{i=1}^{n} z(i) \frac{\delta^{(n)} \;\ln\Xi[\varphi]}
{\prod_{i=1}^{n} \delta z(i)} \right\vert_{\varphi =0} \; .
\end{equation}
It can be shown\cite{Percus} that the 
$\rho_c^{(n)}(1, \ldots,n)$ obey the following hierarchy:
\begin{equation}
\label{hierro}
\frac{\partial \; \rho_c^{(n)}(1, \ldots, n)}{\partial \;\nu}= 
 \int d^3\vec{r}_{n+1}\;\rho_c^{(n+1)}(1, \ldots, n+1) + n \rho_c^{(n)}(1, \ldots,
n) \; .
\end{equation}
Defining now the correlation functions $ h^{(n)}(1, \ldots, n) \equiv
\rho_c^{(n)}(1, \ldots, n)/\rho^n$ we deduce from Eq.\ (\ref{hierro}) important
relations which are repeatedly used in the text, i.e.
\begin{mathletters}
\label{ur}
\begin{eqnarray}
\label{ur1}
\rho^2 \tilde{h}^{(2)}(0) &=& \left(\partial \rho/\partial \nu \right)
-\rho \; ,\\
\label{ur2}
\rho^3\int d^3 \vec{r}_3 \;h^{(3)} (1,2,3)&=&  
\frac{\partial}{\partial \nu} 
\left( \rho^2 h^{(2)}\left( 1,2 \right) \right) 
-2 \rho^2 h^{(2)} (1,2)  \; , \\
\label{ur3}
\rho^3\int d^3 \vec{r}_2 \; d^3 \vec{r}_3 \; h^{(3)} (1,2,3)&=&
 \left(\partial^2 \rho/\partial\nu^2 \right) -3 (\partial \rho /\partial \nu)
+2 \rho \; .
\end{eqnarray} 
\end{mathletters} 
\section{Expression of the $W_N$'s}
\label{appW}
The expressions of $W_2$ and $W_3$ which are needed in the text are obtained
from Eqs.\ (\ref{WN}). The resulting expressions are further simplified by the
use of the relations\ (\ref{ur}) of Appendix\ \ref{appBa}. After
some  elementary  algebra one finds
\begin{mathletters}
\begin{eqnarray}
\label{WW2}
W_2 &=&-\rho^{'}_0 \frac{L^3}{2 a^2} +\frac{\rho^{'}_0}{2a} \int
d^3\vec{r}_1 \; \phi^2(1) -\frac{\rho_0}{4!} \int
d^3\vec{r}_1 \; \phi^4(1)  \nonumber \\
&-&\frac{\rho_0^2}{8} \int
d^3\vec{r}_1 \;d^3\vec{r}_2 \; h_0^{(2)}(1,2) \; \phi^2(1) \phi^2(2) \; , \\
\label{WW3}
W_3 &=& -\rho^{''}_0 \frac{L^3}{6 a^3} + \frac{\rho^{''}_0}{4 a^2}
\int d^3\vec{r}_1 \; \phi^2(1) -\frac{\rho^{'}_0}{4! \;a}
\int d^3\vec{r}_1 \; \phi^4(1) +\frac{\rho_0}{6!}
\int d^3\vec{r}_1 \; \phi^6(1) \nonumber \\
&-& \frac{1}{8a}
\int d^3\vec{r}_1 \;d^3\vec{r}_2 \; \frac{\partial}{\partial \nu_0}
\left(\rho_0^2h_0^{(2)}(1,2)\right) \; \phi^2(1) \phi^2(2) \nonumber \\
&+& \frac{\rho_0^2}{48}
\int d^3\vec{r}_1 \;d^3\vec{r}_2 \; h_0^{(2)}(1,2) \; \phi^2(1) \phi^4(2)
\nonumber \\
&+& \frac{\rho_0^3}{48}
\int d^3\vec{r}_1 \;d^3\vec{r}_2 \; d^3\vec{r}_3 \; 
 h_0^{(3)}(1,2,3) \; \phi^2(1) \phi^2(2) \phi^2(3) \; ,
\end{eqnarray} 
\end{mathletters} 
where $\rho^{'}_0 \text { and } \rho^{''}_0$ denote respectively the first and
second derivatives of the density of the reference hard sphere fluid with
respect to the chemical potential $\nu_0$.
\section{Wick's theorem} 
\label{appWi}
We collect in this Appendix the averages $\langle \ldots \rangle_{X_\tau}$ of
the monomials of $\phi$ which are needed in the text. The following results
have been obtained by making use of Wick's theorem.\cite{Ma,Binney,Parisi}
\begin{eqnarray}
\langle \phi(1)^2 \rangle_{X_\tau} &=& X_\tau(0) \\
\langle \phi(1)^4 \rangle_{X_\tau} &=& 3 X_\tau^2(0) \\
\langle \phi(1)^6 \rangle_{X_\tau} &=& 15X_\tau^3(0) \\
\langle \phi(1)^8 \rangle_{X_\tau} &=& 105X_\tau^4(0) \\
\langle \phi(1)^2 \phi(2)^2  \rangle_{X_\tau} &=& 
2X_\tau^2(12)+X_\tau^2(0) \\
\langle \phi(1)^2 \phi(2)^4  \rangle_{X_\tau} &=& 
12X_\tau^2(12)X_\tau(0)+3X_\tau^3(0) \\
\langle \phi(1)^4 \phi(2)^4  \rangle_{X_\tau}
 &=& 24X_\tau^4(12) + 72X_\tau^2(12)X_\tau^2(0) + 9X_\tau^4(0)\\
\langle \phi(1)^2 \phi(2)^2 \phi(3)^2 \rangle_{X_\tau} &=& X_\tau^3(0) +
8X_\tau(12)X_\tau(23)X_\tau(13) \nonumber \\
&+& 2 X_\tau(0) [ X_\tau^2(12) + X_\tau^2(23)+X_\tau^2(13)] \\
\langle \phi(1)^4 \phi(2)^2 \phi(3)^2 \rangle_{X_\tau} &=& 3X_\tau^4(0)+
48X_\tau(12)X_\tau(23)X_\tau(13)X_\tau(0)+ 24 X_\tau^2(12)X_\tau^2(13)
\nonumber 
\\ &+&  12 X_\tau^2(0) [X_\tau^2(12)+ X_\tau^2(13) ] 
+6 X_\tau^2(23)X_\tau^2(0)   \\
\langle \phi(1)^2 \phi(2)^2 \phi(3)^2  \phi(4)^2
\rangle_{X_\tau}&=&
X_\tau^4(0)  \nonumber \\
&+& 2 X_\tau^2(0) [ X_\tau^2(12)+X_\tau^2(13)+X_\tau^2(14)+
X_\tau^2(23)+X_\tau^2(24)+ X_\tau^2(34) ] \nonumber \\
&+& 8 X_\tau(0) [ X_\tau(12) X_\tau(23) X_\tau(31) + X_\tau(12)
X_\tau(24) X_\tau(41) \nonumber \\
&+& X_\tau(13) X_\tau(34) X_\tau(41) +
X_\tau(23) X_\tau(34) X_\tau(42) ] \nonumber \\
&+& 4[ X_\tau^2(12)X_\tau^2(34) +  X_\tau^2(13)X_\tau^2(24) +
X_\tau^2(14)X_\tau^2(23)] \nonumber \\
&+&16[X_\tau(12)X_\tau(23)X_\tau(34)X_\tau(41) +
X_\tau(12)X_\tau(24)X_\tau(43)X_\tau(31) + \nonumber \\
&+&X_\tau(13)X_\tau(32)X_\tau(24)X_\tau(41)] 
\end{eqnarray}

\section{Computation of $\langle W_2^2\rangle_{X_{\tau,c}}$} 
\label{appW22}
The cumulant $\langle W_2^2\rangle_{X_{\tau,c}}$ is easily obtained from\
(\ref{WW2}) and reads
\begin{mathletters}
\begin{eqnarray}
\langle W_2^2\rangle_{X_{\tau,c}} &=& \sum_{i=1}^{6} X^{i} \; ,\\ 
X^1 &=& \frac{\rho_0^{'2}}{4a^2} \int d^3 \vec{r}_1 d^3\vec{r}_2 \; \left
( \langle
\phi^2(1) \phi^2(2) \rangle_{X_{\tau}} -  \langle \phi^2(1)
\rangle_{X_{\tau}} \langle \phi^2(2)
\rangle_{X_{\tau}}\right) \; , \\
X^2 &=& \frac{\rho_0^{2}}{(4!)^2} \int d^3 \vec{r}_1 d^3\vec{r}_2 \; \left
( \langle
\phi^4(1) \phi^4(2) \rangle_{X_{\tau}} -  \langle \phi^4(1)
\rangle_{X_{\tau}} \langle \phi^4(2)
\rangle_{X_{\tau}}\right) \; , \\
X^3 &=& \frac{\rho_0^4}{64} \int d^3 \vec{r}_1 d^3 \vec{r}_2  d^3\vec{r}_3
d^3\vec{r}_4\; 
h_0^{(2)}(1,2) h_0^{(2)}(3,4)
\left(\langle \phi^2(1) \phi^2(2) \phi^2(3) \phi^2(4) \rangle_{X_{\tau}} \right.
\nonumber \\
& -& \left.
\langle \phi^2(1) \phi^2(2)\rangle_{X_{\tau}} 
\langle \phi^2(3) \phi^2(4)\rangle_{X_{\tau}} \right) \; , \nonumber \\
X^4 &=& -\frac{\rho_0 \rho_0^{'}}{4!a}
\int d^3 \vec{r}_1 d^3\vec{r}_2 \;
\left
( \langle
\phi^2(1) \phi^4(2) \rangle_{X_{\tau}} -  \langle \phi^2(1)
\rangle_{X_{\tau}} \langle \phi^4(2)
\rangle_{X_{\tau}}\right) \; , \\
X^5 &=& -\frac{\rho_0^{'} \rho_0^2}{8a}
\int d^3 \vec{r}_1 d^3 \vec{r}_2  d^3\vec{r}_3\; h_0^{(2)}(2,3)
\left(\langle \phi^2(1) \phi^2(2) \phi^2(3) \rangle_{X_{\tau}}-
\langle \phi^2(1) \rangle_{X_{\tau}} \langle \phi^2(2) \phi^2(3)
 \rangle_{X_{\tau}} \right) \\
 X^6 &=&
 \frac{ \rho_0^3}{4\times 4!}
\int d^3 \vec{r}_1 d^3 \vec{r}_2  d^3\vec{r}_3\; h_0^{(2)}(2,3)
\left(\langle \phi^4(1) \phi^2(2) \phi^2(3) \rangle_{X_{\tau}}-
\langle \phi^4(1) \rangle_{X_{\tau}} \langle \phi^2(2) \phi^2(3)
 \rangle_{X_{\tau}} \right) \; .
\end{eqnarray}
\end{mathletters}
Each of the $X^i$ are now computed by making use of Wick's theorem. It follows
from\ (C1) and (C5) that
\begin{equation}
L^{-3} X^1 = \frac{\rho_0^{'2}}{2a^2} I^{(2)} 
\; ,
\end{equation}
where 
\begin{equation}
I^{(2)} = \int d^3 \vec{r} X_{\tau}^2(r) \; .
\end{equation} 
Similarly, making use of Eq.\ (C2) and (C7) one finds for $X^2$
\begin{equation}
L^{-3} X^2 = \frac{\rho_0^2 X_{\tau}^2(0)}{8} I^{(2)} + \frac{\rho_0^2}{24}
I^{(4)} \; ,
\end{equation}
with
\begin{equation}
I^{(4)} = \int d^3 \vec{r} X_{\tau}^4(r) \; .
\end{equation}
The term $X^3$ is considerably more complicated. It follows from Eqs.\ (C5) and
(C10) that it can be splitted into four pieces
\begin{equation}
\label{lesX}
X^3 = X^{3(a)} + X^{3(b)} + X^{3(c)} + X^{3(d)}
\end{equation}
We have defined $X^{3(a)}$ as
\begin{eqnarray}
\label{bidu}
X^{3(a)}&=& \frac{\rho_0^4 X_{\tau}^2(0)}{8}
\int d^3\vec{r}_1 d^3 \vec{r}_2  d^3\vec{r}_3
d^3\vec{r}_4\;  h_0^{(2)}(1,2) h_0^{(2)}(3,4) X_{\tau}^2(1,3) \nonumber \\
&=& \frac{\rho_0^4}{8}L^3 \tilde{h}_0^2(0) I^{(2)}  X_{\tau}^2(0) \nonumber \\
&=& \frac{1}{8}L^3 I^{(2)} \left( \rho_0^{'}-\rho_0 \right)^2 
 X_{\tau}^2(0) \; ,
\end{eqnarray}
where we made use of the compressibility relation\ (\ref{compres}) to obtain the
last line of Eq.\ (\ref{bidu}).
 The second contribution  $X^{3(b)}$ in Eq.\ (\ref{lesX}) reads as
\begin{eqnarray}
X^{3(b)}&=& \frac{1}{2} \rho_0^4 X_{\tau}(0) 
\int d^3\vec{r}_1 d^3 \vec{r}_2  d^3\vec{r}_3 d^3\vec{r}_4\;  
h_0^{(2)}(1,2) h_0^{(2)}(3,4)X_{\tau}(1,2) X_{\tau}(2,3)X_{\tau}(3,1) \nonumber
\\
&=& \frac{\rho_0^2}{2} L^3 X_{\tau}(0) \left( \rho_0^{'}-\rho_0 \right)
\int d^3 \vec{r}\; h_0^{(2)}(r)  X_{\tau}(r) X_{\tau}^{*(2)}(r) \; ,
\end{eqnarray}
where we have introduced the auto-convolution
\begin{equation}
 X_{\tau}^{*(2)}(r) \equiv
 \int d^3 \vec{r}^{'} X_{\tau}(\vec{r}^{'}) X_{\tau}(\vec{r}-\vec{r}^{'}) \; .
 \end{equation}
The third contribution  $X^{3(c)}$ is given by
\begin{equation}
X^{3(c)}= \frac{\rho_0^4}{8}
\int d^3\vec{r}_1 d^3 \vec{r}_2  d^3\vec{r}_3 d^3\vec{r}_4\;
h_0^{(2)}(1,2) h_0^{(2)}(3,4)
X_{\tau}^2(1,3) X_{\tau}^2(2,4) \; ,
\end{equation} 
 which can  be recast under the simple form
 \begin{equation}
X^{3(c)}= \frac{\rho_0^4}{8} L^3 
\int d^3 \vec{r} \left( h_0^{(2)}*X_{\tau}^2 \right)^2 \; ,
\end{equation}
where $h_0^{2}*X_{\tau}^2 $ denotes the convolution of the functions
$h_0^{(2)}(r)$ and $X_{\tau}^2(r)$. Finally the last contribution  $X^{3(d)}$  
reads as 
\begin{eqnarray}
X^{3(d)}&=& \frac{\rho_0^4}{4}
\int d^3\vec{r}_1 d^3 \vec{r}_2  d^3\vec{r}_3 d^3\vec{r}_4\;
h_0^{(2)}(1,2) h_0^{(2)}(3,4)
 \left[ X_{\tau}(1,2)X_{\tau}(2,3)X_{\tau}(3,4)X_{\tau}(4,1) \right. 
 \nonumber \\
 &+& \left. X_{\tau}(1,2)X_{\tau}(2,4)X_{\tau}(4,3)X_{\tau}(3,1)
 + X_{\tau}(1,3)X_{\tau}(3,2)X_{\tau}(2,4)X_{\tau}(4,1) \right] \; ,
\end{eqnarray}
which can be rewritten as
\begin{eqnarray}
X^{3(d)}&=& \frac{\rho_0^4}{2} L^3 \int d^3 \vec{r} 
\left( \left( h_0^{(2)} X_{\tau} \right) *  X_{\tau} \right)^2(r) \nonumber \\
&+&
\frac{\rho_0^4}{4}
\int d^3\vec{r}_1 d^3 \vec{r}_2  d^3\vec{r}_3 d^3\vec{r}_4\;
h_0^{(2)}(1,2) h_0^{(2)}(3,4)
 X_{\tau}(1,3)X_{\tau}(3,2)X_{\tau}(2,4)X_{\tau}(4,1) \; ,
\end{eqnarray}
where, once again, we have represented the convolution of the two functions 
$h_0^{(2)}(r) X_{\tau}(r)$ and $X_{\tau}(r)$ by the symbol "$*$".

Let us quote now the expressions of the remaining $X^i$ without further 
comments
\begin{mathletters}
\begin{eqnarray}
L^{-3}X^4 &=&  -\frac{1}{2a} \rho_0 \rho_0^{'} X_{\tau}(0)  I^{(2)} \; ,\\
L^{-3}X^5 &=&  -\frac{1}{2a} \rho_0^{'}  \left(\rho_0^{'}-\rho_0 \right)
X_{\tau}(0) I^{(2)}
-\frac{\rho_0^2 \rho_0^{'}}{a} 
\int d^3 \vec{r} \; h_0^{(2)}(r) X_{\tau}(r) X_{\tau}^{*(2)}(r) \; ,\\
L^{-3}X^6 &=& \frac{\rho_0}{4}  I^{(2)} X_{\tau}^2(0) 
\left(\rho_0^{'}-\rho_0 \right) +
\frac{\rho_0^3}{4} \int d^3 \vec{r} \; h_0^{(2)}(r) 
\left( X_{\tau}^2 \; * \; X_{\tau}^2\right)(r) \nonumber \\
&+& \frac{\rho_0^3}{2} X_{\tau}(0) 
\int d^3 \vec{r} \; h_0^{(2)}(r) X_{\tau}(r) X_{\tau}^{*(2)}(r) \; .
\end{eqnarray}
\end{mathletters}

Collecting the expressions of the $X^i$  one finds finally that
\begin{mathletters}
\begin{eqnarray}
\label{uo}
L^{-3} \langle W_2^2 \rangle_{X_{\tau,c}} &=& \sum_{i=1}^7 \omega_i\; , \\
w_1 &=& A_1(a) I^{(2)} \\
w_2 &=&
A_2(a) \int d^3 \vec{r} \; h_0^{(2)}(r) X_{\tau}(r) X_{\tau}^{*(2)}(r) \\
w_3 &=&
 \frac{\rho_0^3}{4}\int d^3 \vec{r} \; h_0^{(2)}(r) \left(X_{\tau}^2 \; *
 \;  X_{\tau}^2 \right)(r)  \\
w_4&=&\frac{\rho_0^4}{8}
\int d^3 \vec{r} \; \left( h_0^{(2)}\; * \; X_{\tau}^2 \right)^2(r)
\\
w_5&=& \frac{\rho_0^4}{2} 
\int d^3 \vec{r} \; \left( \left(h_0^{(2)}X_{\tau}\right)
\; * \; X_{\tau} \right)^2(r) \\ 
w_6 &=&
 \frac{\rho_0^4}{4L^3}\int d^3 \vec{r}_1 d^3 \vec{r}_2  d^3 \vec{r}_3 d^3 \vec{r}_4 
 \; h_0^{(2)}(1,2)h_0^{(2)}(3,4)X_{\tau}(1,3) X_{\tau}(3,2)
X_{\tau}(2,4)X_{\tau}(4,1) \\
w_7 &=&  \frac{\rho_0^2}{24} I^{(4)} \; ,
\end{eqnarray}
\end{mathletters}
with complicated expressions for $A_1(a)$ and $A_2(a)$. However both functions
are regular  in the limit $a \to 0$; more precisely one
has
\begin{mathletters}
\begin{eqnarray}
\label{A1A2}
A_1(a)&=& \frac{1}{8} \rho_0^{'2}\kappa_0^2 + {\cal O}(a) \; , \\
A_2(a)&=& -\frac{1}{2} \rho_0^{2}\rho_0^{'}\kappa_0  + {\cal O}(a) \; .
\end{eqnarray}
\end{mathletters}
We now proceeds in two steps. We first examine the limit $a \to 0$ of each of
the term in the r.h.s of Eq.\ (\ref{uo}) and then study the limit $\gamma \to 0$
of the resulting expression
to decide whether or not it  contributes to $\gamma^4 L^{-3}
\langle W_2^2 \rangle_{X_{\tau}}$ at order ${\cal O}(\gamma^{7/2})$. In other
words, we seek the terms of $ \lim_{a\to 0}
\langle W_2^2 \rangle_{X_{\tau}}$ which are singular in the limit
 $\gamma \to 0$ and discard the regular ones.

Since in the limit $a \to 0$ the potential $X_{\tau}(r)$ tends to the
Yukawa potential $y(r)\equiv \exp(-\kappa_0r)/r$ one has, taking into account
Eq.\ (\ref{A1A2}) 

\begin{equation}
\lim_{a\to 0} w_1=\frac{\rho_0^{'}}{8} \kappa_0^2 
\int d^3 \vec{r} \frac{\exp(-2 \kappa_0 r)}{r^2} \;
 =\frac{\pi \rho_0^{'}}{4}\kappa_0
= {\cal O}(\gamma^{1/2}) \; ,
\end{equation}
and therefore the term $w_1$ can be discarded.
Haga\cite{Haga} has shown that $y^{*(2)}(r)= 2 \pi \exp(-\kappa_0 r)/\kappa_0$ 
from which it follows that 
\begin{equation}
\label{w2}
\lim_{a\to 0} w_2 = -\pi \rho_0^2 \rho_0^{'} \int d^3 \vec{r}
\frac{h_0^{(2)}(r)}{r} + {\cal O}(\gamma^{1/2}) \; .
\end{equation}
Since $h_0^{(2)}(r)$ is a short range function of $r$ the integral in Eq.\
(\ref{w2}) converges and the term $w_2$ can also be discarded.

In the limit $a \to 0$ the third term $w_3$ reads
\begin{equation}
\label{w3}
\lim_{a\to 0} w_3= \frac{\rho_0^3}{4}
\int d^3 \vec{r} h_0^{(2)}(r) \left( y^{*(2)} \; * \; y^{*(2)} \right) (r) \; .
\end{equation} 
It turns out that the function $K(r)=y^{*(2)} \; *\; y^{*(2)} (r)$ has also been
studied by Haga\cite{Haga} who proved that $K(r)=L(r)/r$ where $L(r)$ is an entire function
of $\kappa_0 r$, from which we conclude that $\lim_{a\to 0} w_3$ is regular in
the limit $\gamma \to 0$ and can thus be safely discarded.

In order to study $\lim_{a\to 0} w_4$ we rewrite the integral in Fourier space
and obtain
\begin{equation}
\label{w4}
\lim_{a\to 0} w_4= \frac{\rho_0^4}{8} 
\int \frac{d^3 \vec{q}}{(2 \pi)^3} \left[\tilde{h}_0^{(2)}(q)\right]^2
\left[TF(y^2)(q)\right]^2 \; .
\end{equation}
The Fourier transform of $y^2(r)$ has a simple analytical expression, also given
by Haga\cite{Haga}, i.e.
\begin{equation}
TF(y^2)(q)=\frac{4\pi}{q} \arctan \frac {q}{2 \kappa_0} \; ,
\end{equation}
and therefore 
\begin{equation}
\label{w44}
\lim_{\gamma \to 0} \lim_{a\to 0} w_4=\frac{\pi^4 \rho_0^4}{2}
 \int \frac{d^3 \vec{q}}{(2 \pi)^3} \frac{\left[\tilde{h}_0^{(2)}(q)\right]^2}
{q^2} \; .
\end{equation}
Since the integral\ (\ref{w44}) is convergent the term $\omega_4$ can, as the
previous ones, be discarded.

In order to  study the limit $a \to 0$ of the term $w_5$ we have to consider 
the integral
\begin{equation}
\label{w5}
\int d^3\vec{r} \left[ h_0^{(2)}(r) y(r) \; * \; y(r) \right]^2 \; .
\end{equation}
If we interpret $ z(r) \equiv h_0^{(2)}(r) y(r)$ as a spherically symmetric
distribution of Yukawa
charges, then $v(r)=z*y(r)$ is the potential created by this distribution.
Obviously $v(r) \sim \exp(-\kappa_0 r)/r$ for $r \to \infty$. Therefore the
integrand of eq.\ (\ref{w5}) behaves as $\exp(-2\kappa_0 r)/r^2$ for $r \to
\infty$. The integral\ (\ref{w5}) diverges thus as $1/\kappa_0$ in the limit
$\gamma \to 0$. It follows from this remark that
\begin{equation}
\label{w55}
\lim_{a\to 0} \gamma^4 w_5 = \frac{(\rho_0 \gamma)^{7/2}}{4\pi^{1/2}}
\lim_{\kappa_0 \to 0} 
\int d^3\vec{r}_1 d^3\vec{r}_2 d^3\vec{r}_3 \; h_0^{(2)}(1,2) h_0^{(2)}(3,4)
\frac{\kappa_0 \exp(-\kappa_0(r_{12}+ r_{23} +r_{34}+ r_{41}))}{r_{12} r_{23}
r_{34} r_{41}} \; .
\end{equation}
We come now to the term $w_6$. Taking advantage of the translational invariance
we have
\begin{equation}
\label{w6} 
\lim_{\kappa_0 \to 0} \lim_{a\to 0} w_6 =
\frac{\rho_0^4}{4}
\int d^3 \vec{r}_{12} d^3 \vec{r}_{13} d^3 \vec{r}_{14} \; 
h_0^{(2)} (r_{12}) h_0^{(2)} (r_{34}) \frac{1}{r_{13}} \frac{1}{r_{32}}
\frac{1}{r_{24}} \frac{1}{r_{14}} \;,
\end{equation}
which is a convergent integral due to the short range of the pair correlation
$h_0^{(2)}(r)$. Therefore $w_6$ does not contribute to $\omega_{RPM}$ at order
$\gamma^{7/2}$.

 The last term to examine is $w_7$ which is a tricky one, since
the integral $I^{(4)}$ diverges at short distances in the limit $a \to 0$. In
order to study carefully this divergence we insert in $I^{4}$ the expression\
(\ref{X(r)}) of $X_{\tau}(r)$ which yields
\begin{equation}
\lim_{a \to 0} I^{(4)} = \frac{272}{35} \frac{\pi}{a} +
16 \kappa_0^2 \left(\gamma_E -1 +\ln \left(a\kappa_0 \right)\right)
 +{\cal O}(a) \; ,
\end{equation}
where $\gamma_E$ is Euler's constant. Note that the divergence of $\gamma^4 w_7$
occurs at order $\gamma^4$ and that we do not need to take it into account at
the order of our calculation (i.e. $\gamma^{7/2}$).
 However we have checked that this divergence for
$a \to 0$ originates from the fact that, till now, we have ignored
contributions coming from the cumulant $\gamma^4 \langle W_4
\rangle_{X_{\tau}}$. 
Taking carefully  account  of one of  the terms of this cumulant leads to
consider the integral
\begin{equation}
I^{'(4)} = \int d^3 \vec{r}\; \frac{\exp(-4 \kappa_0 r)}{r^4} g_0^{(2)} (r) 
+{\cal O}(\gamma a)\; 
\end{equation}
instead of $I^{4}$.
Recall that $g_0^{(2)} (r)\equiv h_0^{(2)} (r) + 1$ 
is zero in the core which makes
$I^{'(4)}$ convergent in the limit $a \to 0$ and $\lim_{\gamma \to 0} \lim_{a
\to 0} I^{'(4)}$ perfectly well defined. However the corresponding term 
$\gamma^4 \omega_7^{'}$ contributes to
$\omega_{RPM}$ at order $\gamma^4$ and can thus be discarded.

The conclusion of this lengthy discussion is that 
\begin{eqnarray}
\gamma^4 \langle W_2^2\rangle_{X_{\tau,c}}L^{-3}&=&
\frac{(\rho_0 \gamma)^{7/2}}{4\pi^{1/2}}
\lim_{\kappa_0 \to 0} 
\int d^3\vec{r}_1 d^3\vec{r}_2 d^3\vec{r}_3 \; h_0^{(2)}(1,2) h_0^{(2)}(3,4)
\times \nonumber \\
&\times& \frac{\kappa_0 \exp(-\kappa_0(r_{12}+ r_{23} +r_{34}+ r_{41}))}{r_{12} r_{23}
r_{34} r_{41}} +{\cal O}(\gamma^4)\; .
\end{eqnarray}


\begin{references}
\bibitem{I} previous paper.
\bibitem{Kac} M. Kac, {\em Phys. Fluids} {\bf 2}: 8 (1959).
\bibitem{Hubbard} J. Hubbard, {\em Phys. Rev. Lett.} {\bf 3}: 77 (1954).
\bibitem{Hubbard2} J. Hubbard and P. Shofield, {\em Phys. Lett.} {\bf 77}: 245 (1972).
\bibitem{Stratonovich} R. L. Stratonovich, {\em Sov. Phys. Solid State}
{\bf2}: 1824 (1958). 
\bibitem{Siegert} A. J. F. Siegert, {\em Physica} {\bf 26}: 530 (1960).
\bibitem{Edwards} S.F. Edwards, {\em Phil. Mag.} {\bf 4}: 1171 (1959). 
\bibitem{Samuel} S. Samuel, {\em Phys. Rev. D} {\bf 18}: 1916 (1978). 
\bibitem{Martin} D. C. Brydges and Ph. Martin, {\em J. Stat. Phys.} {\bf 96}:
1163 (1999).
\bibitem{Brillantov} N. V. Brillantov, {\em Phys. Rev. E} {\bf 58}: 2628 (1998).
\bibitem{Orland} R. R. Netz and  H. Orland, {\em Eur. Phys. J. E} {\bf 1}: 67 (2000).
\bibitem{Mayer} J. E. Mayer, {\em J. Chem. Phys.} {\em 18}: 1426 (1950).
\bibitem{Haga} E. Haga, {\em J. Phys. Soc. Japan} {\bf 8}: 714 (1953). 
\bibitem{Fisher} S. Bekiranov and M. E. Fisher, {\em Phys. Rev. E} {\bf 59}:
 492 (1999). 
\bibitem{Percus} {\em see e.g.} J. K. Percus in {\em The equilibrium theory of
classical fluids} edited by H. L. Frisch and J. L. Lebowitz (W. A. Benjamin,
New York Amsterdam, 1964).     
\bibitem{Stell} G. R. Stell and J. Lebowitz, {\em J. Chem. Phys.} {\bf 49}:
3706 (1968). 
\bibitem{Stell2} J. L. Lebowitz, G. Stell, and S. Baer, {\em J. Math. Phys.}
 {\bf 6}: 1282 (1965).
\bibitem{Hemmer} P. C. Hemmer, {\em J. Math. Phys.} {\bf 5}: 75 (1964).
\bibitem{Hauge} E. H. Hauge, {\em J. Chem. Phys.} {\bf 44}: 2249 (1966).
\bibitem{Ma} S. K. Ma, {\em Modern theory of critical phenomena}
(Frontiers in Physics, {\bf 46}).
\bibitem{Parisi} G. Parisi, {\em Statistical Field Theory } (Addison Wesley,
1988).
\bibitem{Binney} J. J. Binney, N. J. Dowrick and A. J. Fisher, M. E. J. Newman,
{\em The  Theory of Critical Phenomena} (Clarendon Press, Oxford, 1992). 
\bibitem{Debye} P. Debye and E. H\"uckel, {\em Physik. Z.} {\bf 24}: 185 (1923).
\bibitem{Hansen} J. P. Hansen and I. Mc Donald, {\em Theory of Simple
Liquids} (Academic, New York, 1986).
\bibitem{Fisher2} M. E. Fisher, {\em J. Stat. Phys.} {\bf 75}: 1 (1994). 
\end{references}
\end{document}